\documentclass[3p, twocolumn]{elsarticle}

\usepackage{lineno,hyperref}
\modulolinenumbers[5]

\usepackage{epsfig}
\usepackage{amsfonts}
\usepackage{amsmath}
\usepackage{mathrsfs, amssymb, amsfonts, amsmath, mathtools}
\usepackage{bm}
\usepackage{float}
\usepackage[caption=false]{subfig}
\usepackage[utf8]{inputenc}
\newcommand{\bra}[1]{\langle #1|}
\newcommand{\ket}[1]{|#1\rangle}

\newcommand{\beq}{\begin{equation}}
\newcommand{\eeq}{\end{equation}}
\newcommand{\bea}{\begin{align}}
\newcommand{\eea}{\end{align}}
\newcommand{\nn}{\nonumber}

\begin{document}

\bibliographystyle{elsarticle-num}

\begin{frontmatter}

\title{Flux sensitivity of quantum spin Hall rings}


\author[wu]{F. Cr\'epin }
    \ead{francois.crepin@physik.uni-wuerzburg.de}
\author[wu]{B. Trauzettel }

\address[wu]{Institute for Theoretical Physics and Astrophysics,
University of W\"urzburg, 97074 W\"urzburg, Germany}



\begin{abstract}
We analyse the periodicity of persistent currents in quantum spin Hall loops, partly covered with an $s$-wave superconductor, in the presence of a flux tube. Much like in normal (non-helical) metals, the periodicity of the single-particle spectrum goes from $\Phi_0 = h/e$ to $\Phi_0/2$  as the length of the superconductor is increased past the coherence length of the superconductor. We further analyze the periodicity of the persistent current, which is a many-body effect. Interestingly, time reversal symmetry and parity conservation can significantly change the period. We find a $2\Phi_0$-periodic persistent current in two distinct regimes, where one corresponds to a Josephson junction and the other one to an Aharonov-Bohm setup. 
\end{abstract}


\end{frontmatter}


\section{Introduction}
\begin{figure*}
\centering
       \includegraphics[width=12cm,clip]{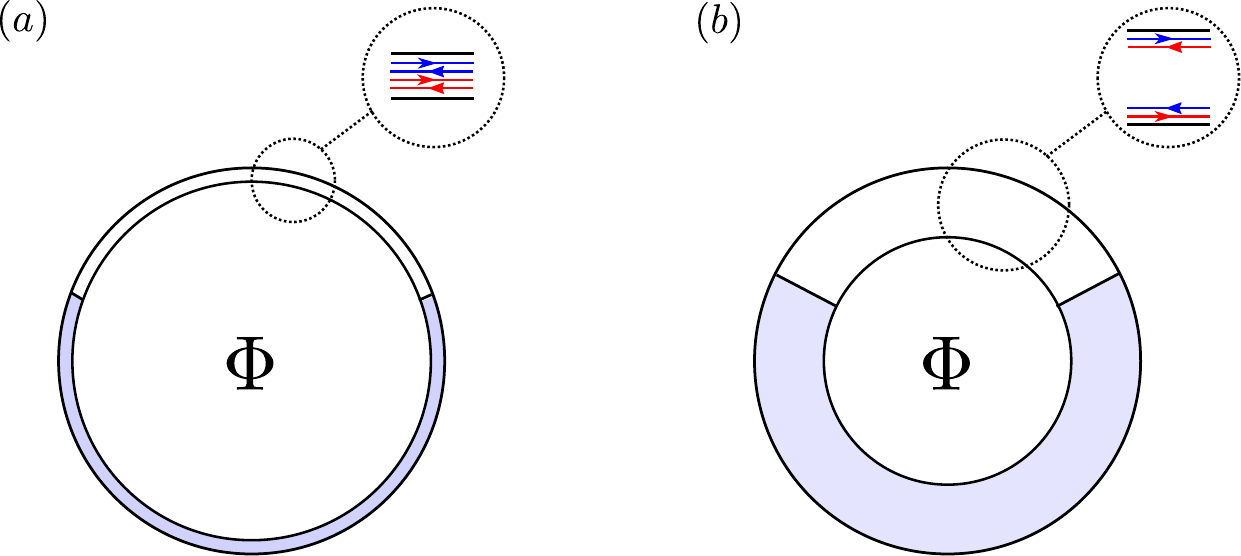}
		 \caption{(a) Original setup as proposed in Ref.~\cite{Klapwijk86}. A piecewise normal and superconducting loop of size $L$ comprising a single conducting channel is threaded by a magnetic flux $\Phi$. Close to the Fermi points, there are four available modes: spin up or spin down right movers (blue solid lines) and spin up or spin down left movers (red solid lines). The normal region has length $d_N$ while the superconducting region has length $d_S$. (b) A quantum spin Hall loop, partly covered by an $s$-wave superconductor (shaded region) and threaded by a flux $\Phi$. Each edge hosts half the degrees of freedom available in setup (a), resulting in two helical channels physically separated by the insulating bulk.}\label{Fig:setup}
\end{figure*}

In a seminal work from 1986, B\"uttiker and Klapwijk discussed the flux sensitivity of a piecewise normal and superconducting metal loop~\cite{Klapwijk86}, as showed in Fig.~\ref{Fig:setup}(a). The model they considered in order to describe such a system -- a single electronic channel with a linearized spectrum -- is as simple as it gets, yet captures the characteristic features associated with persistent currents in mesoscopic loops. Indeed, in the Andreev approximation, and in the low-energy regime, microscopic details of the model hardly matter. As long as the length of the normal region is much larger than the coherence length of the superconductor, the persistent current will have the familiar saw tooth shape, both in the normal and superconducting regime. What changes though between the two regimes is the periodicity of the superconducting current with the applied flux. Following a simple calculation of the excitation spectrum, Büttiker and Klapwijk were able to show how as the length of the superconducting region is progressively increased, the periodicity of the persistent current is halved, going from $\Phi_0 $  to $\Phi_0/2$, with $\Phi_0 = h/e$ the quantum of flux. The almost thirty years since the 1986 paper have seen many exciting discoveries in the field of mesoscopic physics. One of them is the advent of topological insulators~\cite{Kane05, Kane05b, Zhang06b, Konig07}. Of particular interest to us here is the case of quantum spin Hall insulators and their one-dimensional helical edge states, for several reasons. They first offer a new realization of 1D Dirac physics, that goes beyond linearization of quadratic spectra at the Fermi points. Second, helicity, that is the locking of direction of spin with direction of motion, protects transport against time-reversal invariant impurities. More precisely, single-particle elastic backscattering is forbidden by time-reversal symmetry, leading the community in a vast effort to better understand the effect of inelastic scattering in these systems~\cite{Japaridze10,Budich12, Schmidt12, Oreg12, Crepin12, Konig12, Glazman13, Glazman14, Pikulin14, Geissler14, Kainaris14}. Third, the interplay of helicity and superconductivity imposes a constraint on the parity of the number of quasi-particles, or fermion parity (FP), in SNS junctions based on helical liquids, as the one depicted in Fig.~\ref{Fig:setup}(b). This results in a so-called fractional Josephson effect~\cite{Fu09b, Beenakker12, Crepin14a}, with a $\Phi_0$-periodic supercurrent. A few works have already discussed the physics of persistent currents in helical rings, highlighting the effects of magnetic and non-magnetic impurities~\cite{Sticlet13}, or the hybridization between edge states in narrow quantum spin Hall rings~\cite{Michetti11}. In the present paper, we  revisit the analysis of Büttiker and Klapwijk in the context of quantum spin Hall insulators. In particular we analyse the crossover between the normal persistent current and the supercurrent, as the length of the superconducting region is increased, and discuss their periodicity. We argue that a constraint from time reversal symmetry doubles the period in the normal case, as compared to Ref.~\cite{Klapwijk86}. {A similar effect is already known to occur in the superconducting case, and was put forward by Zhang and Kane in Ref.~\cite{Kane14b}.}

The outline of the paper is as follows. In Sec.~\ref{Sec:model}, we start by discussing the model and the various symmetries that constrain the periodicity at a given edge, both in the normal and superconducting regimes, and contrast the results with the case of a non-helical metal. Then, in Sec.~\ref{Sec:total}, we comment on the total persistent current, when both edges are taken into account. Finally, in Sec.~\ref{Sec:conclusions}, we give some conclusions. 

\section{Model and persistent current at a given edge} \label{Sec:model}

\subsection{Normal helical ring}

\subsubsection{Single-particle spectrum}

We are interested in modeling the setup of Fig.~\ref{Fig:setup}(b). To that end, we first restrict our analysis to a single edge, say the outer one. Before following in the footsteps of Ref.~\cite{Klapwijk86} and computing the excitation spectrum in the superconducting case, let us consider first the normal case. We model the outer edge by a 1D segment of size $L$ along the $x$ direction and impose periodic boundary conditions. The flux is included via the minimal substitution of the momentum operator $\hat{p}_x= -i \hbar \partial_x$,
\beq
\hat{p}_x \to \hat{p}_x - q A\; 
\eeq
with $q$ the charge of the particles and $A = \Phi/L$ a vector potential. In the following we take $q = -e$, for electrons. At energies much smaller than the bulk band gap, the helical states are well described by Dirac fermions, with the following single particle Hamiltonian
\beq
\mathcal{H} = v_F (\hat{p}_x +e A) \sigma_3 \;, \label{Eq:H_sp}
\eeq
where $v_F$ is the Fermi velocity and $\sigma_3$ is the usual Pauli matrix, acting on spin space. Right moving electrons have therefore spin up while left moving electrons have spin down (the situation is reversed at the inner edge). The single particle spectrum consists of two branches \beq
\varepsilon_{\pm,n}(\Phi) = \pm \hbar v_F  \frac{2\pi}{L}\left(n + \frac{\Phi}{\Phi_0} \right)\;, \label{Eq:spec_sp}
\eeq
with $\Phi_0 = h/e$. Corresponding eigenstates are of the form
\beq
\phi_{\pm,n}(x) =  \chi_\pm e^{i k_n x}/\sqrt{L}\;,
\eeq
with $\chi_+ = (1,0)^T$, $\chi_- = (0,1)^T$ and the momentum $k_n = 2\pi n /L$ is quantized due to periodic boundary conditions. 

\subsubsection{Excitations}
\label{Sec:excitations}

In the absence of the flux, $\Phi = 0$, we take the chemical potential to be at the Dirac point, that is, all states with negative energy are filled. Given the Hamiltonian of Eq.~\eqref{Eq:H_sp}, nothing indicates that the spectrum is bounded from below, which would mean the ground state energy is infinite. Of course, in a real system, there is a natural cutoff scale, as the spectrum is bounded both from below and above by the bulk bands. However, at this stage one can let the cutoff go to infinity and renormalize the ground state energy without affecting the general physics which is given by the low energy excitations with respect to the Fermi sea. We then define the ground state $\ket{0}_0$ such that
\begin{align}
c^\dagger_{\pm,n}\ket{0}_0 &= 0\;, \quad n\leq 0\;, \\
c_{\pm,n}\ket{0}_0 &= 0\;, \quad n>0\;. 
\end{align}
and we impose that $\ket{0}_0$ has zero energy. The index $0$ here serves as a reminder that the state is defined for $\Phi=0$. There are two types of excitations on top of  $\ket{0}_0$. One can create a particle in the conduction band, for instance $c^\dagger_{\pm,n>0}\ket{0}_0$, or create a hole in the valence band, for instance $c_{\pm,n\leq0}\ket{0}_0$. Importantly, particle and hole excitations are independent. We can also define a many-body Hamiltonian (still for $\Phi = 0$) as
\beq
H = \int_0^L dx \; :\Psi^\dagger(x) \mathcal{H} \Psi(x): \;, \label{Eq:H_mb}
\eeq
with $\Psi(x) = [\psi_+(x), \psi_-(x)]^T$ a quantum field and where $: \ldots :$ indicates normal-ordering, ensuring that indeed $_0\bra{0} H \ket{0}_0 = 0$. Among the many possible excited states, some will play a particular role in the following. These are the states with a finite number of particles, but no particle-hole excitations. They are obtained by filling positive energy states or emptying negative energy states in the following way. The state with $N_{j = \pm}$ particles is defined by
\begin{align}
\ket{N_j}_0 &= \prod_{n=1}^{N_j} c^\dagger_{j,n}\ket{0}_0\;, \quad &\textrm{if} \quad N_j>0\;, \\
\ket{N_j}_0 &= \prod_{n=0}^{N_j-1} c_{j,-n}\ket{0}_0\;, \quad &\textrm{if} \quad N_j<0\;, 
\end{align}
In the following we use the notation $\ket{N_+, N_-}_0$ in order to refer to a state with $N_+$ right movers and $N_-$ left movers. The energy of these many-body states is simply given by~\cite{Vondelft98}
\beq
E(N_+,N_-,\Phi=0) = \hbar v_F \frac{2\pi}{L} \frac{1}{2} \sum_{j=\pm} N_j (N_j+1)\;. 
\eeq
This prompts us to introduce the so-called chiral current operators
\beq
J_\pm(x) = \; :\psi_\pm^\dagger(x) \psi_\pm(x):\;, \label{Eq:chiral_current}
\eeq
as well as the particle current operator
\beq
J(x) = v_F :\Psi^\dagger(x) \sigma_3 \Psi(x):\;. \label{Eq:current}
\eeq
Next, we want to consider the effects of a finite flux $\Phi \neq 0$ on the many-body states. There are several ways to look at the problem, some of which can lead to paradoxical claims, all of which are more or less related to the absence of a lower bound in the spectrum. The first observation is that the flux disturbs the vacuum. Indeed, at the  level of the single-particle spectrum, all states move up and down when $\Phi$ is varied. When one quantum of flux $\Phi_0$ has been threaded, the single-particle spectrum has mapped onto itself, however the many-body state has changed, as an electron-hole pair has been created on top of the vacuum
\beq
\ket{0}_0 \longrightarrow c^\dagger_{+,1} c_{-,0}\ket{0}_0 = \ket{N_+ = 1, N_- = -1}_0\;.
\eeq
Using the definitions of Eqs.~\eqref{Eq:chiral_current} and \eqref{Eq:current}, one finds that this state carries a finite current $2v_F/L$. However, we would like to compute the current and the energy for an arbitrary value of $\Phi$ between 0 and $\Phi_0$. The problem is that these quantities are ill-defined, since our regularization scheme was introduced only for $\Phi = 0$. One way to circumvent this problem is to use perturbation theory. We assume that the field is turned on at some time $t_0$ and compute the current at a later time $t$. We decompose $H$ into 
\beq
H = H_0 + \theta(t-t_0)H_1\;,
\eeq
with 
\begin{align}
H_0 &= v_F \int_0^L dx\;  \Psi^\dagger(x) \hat{p}_x \sigma_3 \Psi(x) \;, \\
H_1 &=  e A \int_0^L dx\;  J(x) \;,
\end{align}
and treat $H_1$ in linear response. From the Kubo formula we have
\begin{align}
\langle J_{\pm}(x,t) \rangle &= \langle J_{\pm}(x,t) \rangle_0 \nn \\ 
& \hspace{-1cm} + \frac{e A}{\hbar} \int_{t_0}^\infty dt_1 \; \int_0^L dx_1\; C^R(x,t,x_1,t_1)\;, \label{Eq:Kubo}
\end{align}
with
\beq
 C^R(x,t,x_1,t_1) = -i \theta(t-t_1) \langle \left[ J_{\pm}(x,t) , J(x_1,t_1)  \right]\rangle_0\;.
\eeq
Recognizing the famous anomalous commutators of bosonization~\cite{Giamarchi-book},
\beq
\langle [J_{\sigma}(q), J_{\sigma'}(-q')] \rangle_0= - \delta_{\sigma,\sigma'}\delta_{q,q'}\frac{\sigma q L}{2\pi}\;,
\eeq
with  $\displaystyle J_{\sigma = \pm}(q) = \int_0^L dx \; e^{-iqx} J_{\sigma}(x) $, we finally arrive at
\beq
\langle J_{\pm}(x,t) \rangle = \frac{1}{L} \left( N_{\pm} \pm \frac{\Phi}{\Phi_0} \right)\;, \quad t>t_0\;. \label{Eq:Jpm}
\eeq
Eq.~\eqref{Eq:Jpm} shows clearly that the number of right (left) movers changes by $+1$ ($-1$) as one quantum of flux is threaded. This is a manifestation of the so-called chiral anomaly of $(1+1)d$ quantum electrodynamics~\cite{GogolinBook}. In the context of condensed matter though, one is expected to provide a physical explanation to resolve the anomaly. One way is to think of an underlying 1D lattice of which the Hamiltonian of Eq.~\eqref{Eq:H_sp} is only an approximation close to the Fermi points. The effect of the flux is then to push all states in one direction in the Brillouin zone, effectively converting one left mover into a right mover. However, for helical liquids, the more appropriate way of thinking is to involve the two edges of the setup, which exchange particles through the bulk as the flux is increased. A more detailed discussion of the persistent current when both edges are involved can be found in Sec.~\ref{Sec:total}. We can also derive the correction to the energy coming from the anomalous current. One straightforward way is to realize that the current flows, in the presence of the induced electric field $E(x,t) = -\partial_t A(x,t)$. This results in an additional energy of 
\begin{align}
E_{W} &= \int_{t_0}^t dt' \int_0^L dx \; (-e) \langle J(x,t') \rangle  E(x,t') \nn \\
& = \hbar v_F \frac{2\pi}{L} (N_{+} -N_{-}) \frac{\Phi }{\Phi_0} + \hbar v_F \frac{2\pi}{L} \left(\frac{\Phi }{\Phi_0} \right)^2\;.
\end{align}
In turn, the ground state energy in the presence of the flux becomes
\begin{align}
E(N_+,N_-,\Phi) &= \hbar v_F \frac{2\pi}{L} \frac{1}{2} \sum_{j=\pm} N_j (N_j+1) \nn \\
& \hspace{-2cm} +\hbar v_F \frac{2\pi}{L} (N_{+} -N_{-}) \frac{\Phi }{\Phi_0} + \hbar v_F \frac{2\pi}{L} \left(\frac{\Phi }{\Phi_0} \right)^2\;. \label{Eq:MB_flux}
\end{align}
The energies of several states are represented in Fig.~\ref{Fig:MB_spectrum}.  With Eq.~\eqref{Eq:MB_flux}, we can recover the usual expression for the persistent current at zero temperature
\beq
I(\Phi) =  \frac{\partial E}{\partial \Phi}\;.\label{Eq:I}
\eeq
The approach we have used here is in our sense physically transparent and does not put too much emphasis on the unphysical infinite spectrum. In regular metals, with e.g a quadratic dispersion, the energy and the current can be computed directly by summing up over the finite number of occupied states below the Fermi energy. In the Dirac case, an analogous approach is available and consists in regularizing the infinite sum with a smooth cutoff, therefore mimicking the bandwidth. A detailed calculation, similar to the one leading to the Casimir effect, is nicely described in Ref.~\cite{Sticlet13}.

\subsubsection{Persistent current}

\label{Sec:Pers}

\begin{figure}[h]
\centering
       \includegraphics[width=6cm,clip]{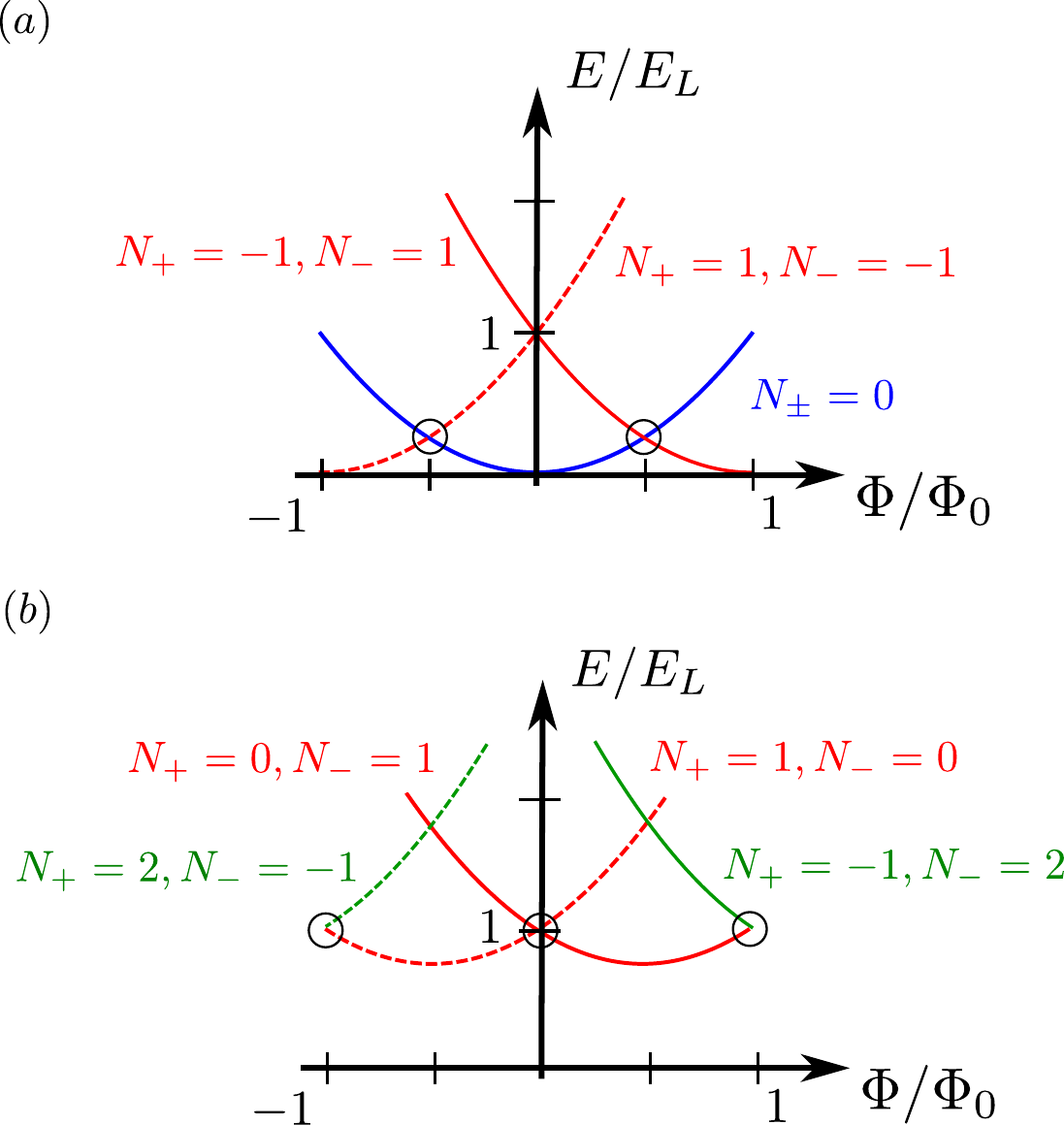}
		 \caption{Plot of the energies of several many-body states $E(N_+,N_-,\Phi)$, with $E_L = h v_F/L$. We have represented the lowest states with (a) $N_++N_- = 0$, (b) $N_++N_- = 1$. }\label{Fig:MB_spectrum}
\end{figure}

\begin{figure}[h]
\centering
       \includegraphics[width=6cm,clip]{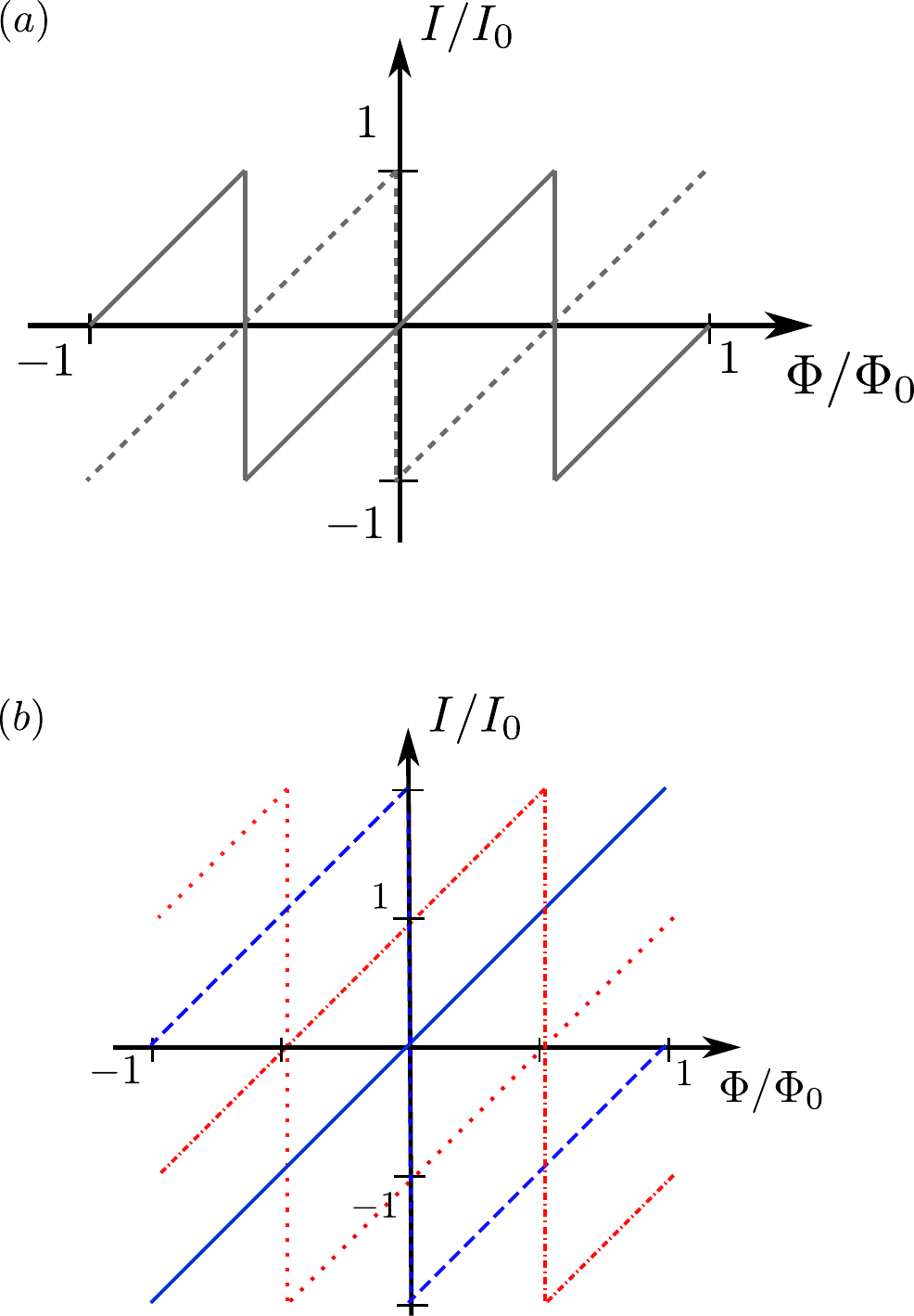}
		 \caption{Plot of the persistent current for different starting points, in units of $I_0 = e v_F/L$. (a) In the normal (non-helical) case there are two inequivalent starting points and the persistent current is $\Phi_0$ periodic. The solid line corresponds to starting with $N_+ = N_- = 0$ at $\Phi =0$ while the dashed line corresponds to starting with either $N_+= 1$ and $N_- = -1$ (positive current) or $N_+= -1$ and $N_- = 1$ (negative current). (b) In the helical case, taking into account the constraint from time reversal symmetry, there are four inequivalent starting points and the persistent current is $2\Phi_0$ periodic. The blue solid line corresponds to $N_+ = N_- = 0$ at $\Phi =0$, the blue dashed line to either $N_+= 1$ and $N_- = -1$ or $N_+= -1$ and $N_- = 1$, the red dashed-dotted line to $N_+= 1$ and $N_- = 0$ and finally the red dotted line to  $N_+= 0$ and $N_- = 1$.  }\label{Fig:current_1}
\end{figure}

Now, the question arises of what happens for large values of the flux. Increasing the flux requires more and more energy, therefore driving the system far away from the ground state. However, for some values of the flux (integer or half-integer multiples of $\Phi_0$), different many-body states become degenerate, as shown in Fig.~\ref{Fig:MB_spectrum}, and the system can in principle use these degeneracies  to relax to the ground state. The consequence of this mechanism is a periodic persistent current. As was emphasized in Ref.~\cite{Klapwijk86}, the microscopic details of the model hardly play a role, and Eqs.~\eqref{Eq:MB_flux} and \eqref{Eq:I} can, at the end of the day, in principle be used both for a normal and a helical channel.  The crucial difference however, will lie in determining which crossings are protected by symmetry. Indeed, in the helical case, the degeneracies circled in Fig.~\ref{Fig:MB_spectrum} are protected by time reversal symmetry. To connect these states, one has to backscatter a right mover with spin up into a left mover with spin down, or vice versa. Barring the presence of magnetic impurities, matrix elements between these states are usually zero, and no gap opens. The other crossings concern states that can be connected by two-particle backscattering events which are not forbidden by time-reversal symmetry. As a consequence, the persistent current in the helical case has twice the period as in the non-helical case described in Ref.~\cite{Klapwijk86}. The system only returns to the ground state after threading $\Phi = 2\Phi_0$. In Fig.~\ref{Fig:current_1}, we give several plots of the persistent current, both in the normal and helical case, for different starting points. The doubling of the periodicity due to the topological protection against single particle backscattering in helical liquids is the main departure from the results of Büttiker and Klapwijk. Our analysis does not however take into account effects of inelastic processes which can lead to transitions between bands~\cite{Buttiker83, Buttiker85, Buttiker85b, Fu09b}{, and provide the system with a way to bypass the topological protection, relax to the ground state and restore the $\Phi_0$ periodicity.  In the case where the length of the ring is small compared to the inelastic mean free path (a few $\mu$m in typical quantum spin Hall insulators such as HgTe/CdTe~\cite{Konig07} or InAs/GaSb~\cite{Knez11}), the energy level broadening is small compared to the level spacing and the band structure given by Eq.~\eqref{Eq:H_sp} is preserved~\cite{Buttiker85b}. \newline
\indent If a measurement of the persistent current (in the normal state regime) were done on time scales larger than the corresponding inelastic scattering time (in the system under consideration) then the protection due to time reversal symmetry would be relaxed. The absence of this constraint would result in a $\Phi_0$ periodicity of the persistent current. For InAs/GaSb quantum wells, with a Fermi velocity of $v_F \simeq 10^4$ m.s$^{-1}$~\cite{Knez14}, the inelastic scattering time is estimated to be of the order of 0.1ns.}

\subsection{Supercurrent in SNS junctions}

\subsubsection{Excitation spectrum}

\begin{figure}
\centering
       \includegraphics[width=7cm,clip]{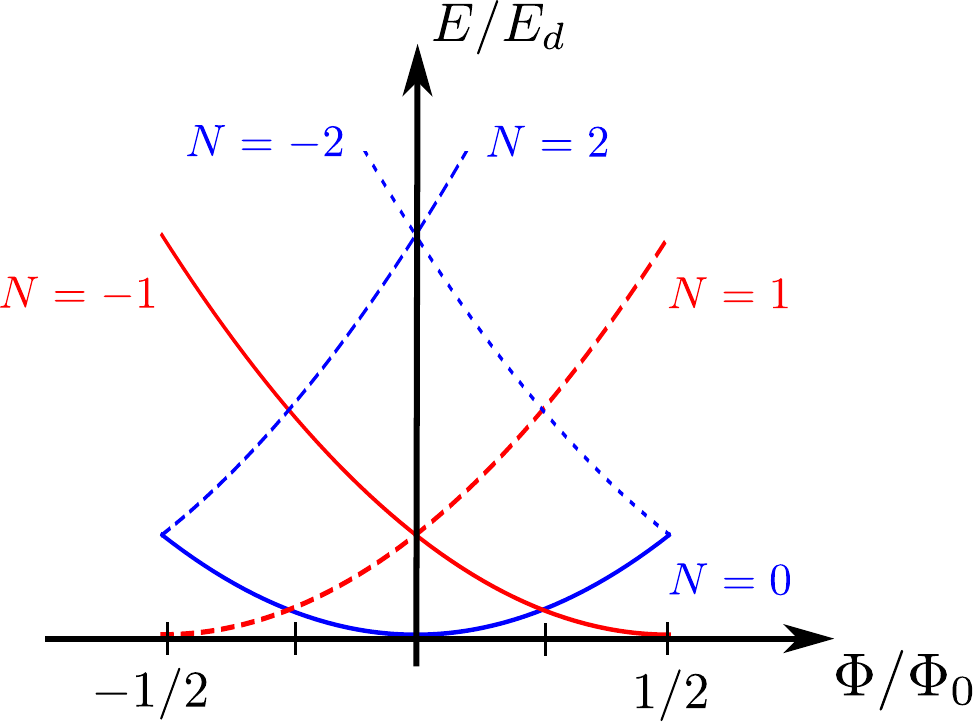}
		 \caption{Plot of the energies of several many-body states $E(N,\Phi)$, with $E_d = h v_F/(2d_N)$. We have represented the lowest states with $N=0,\pm 1,\pm 2$. }\label{Fig:BdG_spectrum}
\end{figure}

We now turn to the case where part of the topological insulator is covered with an ordinary $s$-wave superconductor which induces pairing by proximity effect. We still focus for now on the outer ring of size $L$ and assume that pairing is induced on a length $d_S$. The remaining normal region has length $d_N$. The crucial difference with the previous case is that, in a mean-field description of superconductivity, electrons and holes are not independent excitations anymore. The system is well described by the following many-body Hamiltonian
\beq
H = \frac{1}{2}\int_0^L dx \Psi^\dagger(x) \mathcal{H}_{\textrm{BdG}}(x) \Psi(x)\;,
\eeq
where now $\Psi= [\psi_+, \psi_-, \psi_-^\dagger, -\psi_+^\dagger]^T$ and
\beq
\mathcal{H}_{\textrm{BdG}}(x) =  \begin{pmatrix}
v_F(\hat{p}_x+e A) \sigma_3 & \Delta(x) e^{- 2 i e A x/\hbar} \\
\Delta(x) e^{2 i e A x/\hbar} & -v_F(\hat{p}_x-e A) \sigma_3
\end{pmatrix}\;, \label{Eq:HBdG_1}
\eeq
is the Bogoliubov-de Gennes Hamiltonian, with $\Delta(x) = \Delta_0 \theta(x)\theta(d_S-x)$, and $\theta(x)$ the Heaviside function. At the heart of Ref.~\cite{Klapwijk86}, is the computation of the spectrum for arbitrary values of $d_N$ and $d_S$. The spectrum has two branches given by the eigenvalues of $\mathcal{H}_{\textrm{BdG}}$ in  Eq.~\eqref{Eq:HBdG_1}, and its periodicity with the flux interpolates between $\Phi_0$ and $\Phi_0/2$ as the length of the superconducting region is increased. In the limit $d_S, d_N \gg \xi$ with $\xi = \hbar v_F /\Delta_0$ the coherence length of the superconductor, the single-particle spectrum reads
\beq
\varepsilon_{\pm,n}(\Phi) = \hbar v_F \frac{\pi}{d_N}\left(n +\frac{1}{2}\pm 2\frac{\Phi}{\Phi_0}\right)\;.
\eeq
Incidentally, such a spectrum is that of 1D Dirac fermions on a ring of size $2d_N$, with twisted boundary conditions~\cite{Maslov96, Crepin14a}. Building on Eq.~\eqref{Eq:MB_flux}, we could then write the following expression for the energy of a given many-body state
\beq
E(N_+, N_-, \Phi) = \frac{1}{2} \sum_{j=\pm} \hbar v_F\frac{\pi}{2d_N}\left( N_j  \pm 2 \frac{\Phi}{\Phi_0} \right)^2\;.
\eeq
Note that the expression is more compact than in Eq.~\eqref{Eq:MB_flux} because in the absence of a flux the two consecutive Andreev reflections at the NS interfaces impose effective antiperiodic boundary conditions. However, due to the built-in particle-hole symmetry of the BdG Hamiltonian, "right" and "left" movers are not independent excitations anymore, and only one fermion number is needed to describe excitations in the junction~\cite{Maslov96, Crepin14a}. More precisely we have $N_+ = -N_- = N$.  The final expression for the energy of a state with $N$ quasi-particles is then
\beq
E(N,\Phi) = \hbar v_F\frac{\pi}{2d_N}\left( N  + 2 \frac{\Phi}{\Phi_0} \right)^2\;.
\eeq
We see that as the flux is increased by half a quantum of flux the number of quasi-particles in the junction is changed by one. This is another manifestation of the chiral anomaly, now for Majorana fermions, and is sometimes dubbed fermion parity anomaly in the context of topological superconductors~\cite{Beenakker12}. The energies of several states are represented in Fig.~\ref{Fig:BdG_spectrum}. A more precise, microscopic calculation along the lines of Sec.~\ref{Sec:excitations} would lead to the same expression. Note that in the non-helical case, all many-body states are twice degenerate. Finally, the persistent current at zero temperature is still given by 
\beq
I(\Phi) =  \frac{\partial E}{\partial \Phi}\;.\label{Eq:I_BdG}
\eeq

\subsubsection{Persistent current}

In the helical case, the crossings at $\Phi = \pm\Phi_0/4$ between the states with $N=0$ and $N=\mp 1$ are protected by fermion parity conservation~\cite{Fu09b, Beenakker12}. Crossing these values of the flux, the system cannot relax to the ground-state, and as a consequence, the persistent current is $\Phi_0$ periodic instead of $\Phi_0/2$ periodic in the case of a non-helical junction. Furthermore, it has recently been put forward that time-reversal symmetry might double again the period~\cite{Kane14b}. Even though the analysis was originally done in the short junction limit, the simple arguments exposed in Sec.~\ref{Sec:Pers} together with the spectrum of Fig.~\ref{Fig:BdG_spectrum} can be used to understand the long junction regime. The idea is that the crossing at $\Phi = 0$ between the $N =\pm 1$ states is protected by time-reversal symmetry while the crossing between the $N = \pm 2$ states is not and could be opened by two-particle backscattering processes arising from Coulomb interactions. This would, once again, result in a $2\Phi_0$ periodic persistent current.

\section{Total current including both edges} \label{Sec:total}

\begin{table}
  \resizebox{0.47\textwidth}{!}{  \begin{tabular}{ | c  || c | c || c | c |}
    \hline
  System  & \multicolumn{2}{c || }{Spinful} &  \multicolumn{2}{c | }{Helical} \\ \hline \hline
  Regime  & $d_S < \xi$ & $d_S \gg \xi$ & $d_S < \xi$ & $d_S \gg \xi$ \\ \hline
  Spectrum & $\Phi_0$ & $\Phi_0/2$ & $\Phi_0 $ & $\Phi_0/2$ \\ \hline
  Current & $\Phi_0$ & $\Phi_0/2$ & $\alpha \Phi_0$ & $ \beta \Phi_0 $ \\ \hline
    \end{tabular}}
        \caption{Periodicity of the spectrum and of the persistent current in the spinful and helical case, in the normal regime (length of the superconductor $d_S$ much smaller than the coherence length $\xi$) and in the superconducting regime (length of the superconductor $d_S$ much bigger than the coherence length $\xi$). In the helical case, $\alpha = 2$ if the constraint of time reversal symmetry is taken into account, $\alpha = 1$ if it is relaxed. Similarly, $\beta = 2$ if both the constraints from time reversal symmetry and fermion parity conservation are taken into account, $\beta = 1$ if time reversal symmetry is relaxed and finally $\beta = 1/2$ if fermion parity conservation is further relaxed. } \label{T:Period}
\end{table}

We have seen that in quantum spin Hall systems helicity and time-reversal symmetry impose new constraints on the many-body states, that can change the periodicity of the persistent current. In Table~\ref{T:Period}, we summarize the situation. Of course, such an experiment with a flux threading would involve both edges of the quantum spin Hall insulator. The question then arises what is the behavior of the total persistent current given by the sum of the persistent currents at each edge. In the obvious case where the width of the topological insulator becomes small, then time reversal symmetry would not protect the system against elastic single-particle backscattering (as an electron could change direction without flipping spin, simply by tunneling from one edge to the other one) and the total persistent current would recover the periodicity of its counterpart in the spinful metal. In reality, tunneling between edges can never be really suppressed and gaps will always open in the spectrum. What matters then is the time scale set by such gaps compared to the rate at which the flux is varied. Furthermore, even in the ideal limit of a decoupling between edges, they are never truly independent. Indeed, they are both needed in order to explain physically the chiral anomaly, described in Sec.~\ref{Sec:excitations}. In the normal regime, we have seen that as one increase the flux by one quantum of flux, a particle-hole excitation is created at the outer edge. Although no charge has been created, a left mover has been converted into a right mover, or, using our convention for helicity, a spin down has been converted into a spin up. This is only possible because, similarly to the Laughlin argument for the quantum Hall effect, the opposite process happens simultaneously at the outer edge and the whole system acts as a spin pump~\cite{Fu09b, Fu06}.

\section{Conclusions} \label{Sec:conclusions}

Following the ideas outlined by Büttiker and Klapwijk in Ref.~\cite{Klapwijk86}, we have revisited the question of persistent currents in quantum spin Hall rings, with or without proximity induced superconductivity. We have found that contrary to a spinful metal, helicity together with time reversal symmetry puts additional constraints on the transitions between degenerate states and, as a result, can double the periodicity of the persistent current. 

\section{Acknowledgments} \label{Sec:ack}
We enjoyed physics discussions with Markus Büttiker at many conferences and mutual visits.  Financial support by the DFG (through SFB 1170 "TocoTronics") is gratefully acknowledged.


\section*{References}
\bibliography{Top_ins_wurzburg}

\begin{thebibliography}{10}
\expandafter\ifx\csname url\endcsname\relax
  \def\url#1{\texttt{#1}}\fi
\expandafter\ifx\csname urlprefix\endcsname\relax\def\urlprefix{URL }\fi
\expandafter\ifx\csname href\endcsname\relax
  \def\href#1#2{#2} \def\path#1{#1}\fi

\bibitem{Klapwijk86}
M.~B\"uttiker, T.~M. Klapwijk, Flux sensitivity of a piecewise normal and
  superconducting metal loop, Phys. Rev. B 33 (1986) 5114--5117.

\bibitem{Kane05}
C.~L. Kane, E.~J. Mele, $z2$ topological order and the quantum spin hall
  effect, Phys. Rev. Lett. 95~(14) (2005) 146802.

\bibitem{Kane05b}
C.~L. Kane, E.~J. Mele, Quantum spin hall effect in graphene, Phys. Rev. Lett.
  95 (2005) 226801.

\bibitem{Zhang06b}
B.~A. Bernevig, T.~L. Hughes, S.-C. Zhang, Quantum spin hall effect and
  topological phase transition in hgte quantum wells, Science 314~(5806) (2006)
  1757--1761.

\bibitem{Konig07}
M.~K\"onig, S.~Wiedmann, C.~Br\"une, A.~Roth, H.~Buhmann, L.~W. Molenkamp,
  X.-L. Qi, S.-C. Zhang, Quantum spin hall insulator state in hgte quantum
  wells, Science 318~(5851) (2007) 766--770.

\bibitem{Japaridze10}
A.~Str\"om, H.~Johannesson, G.~I. Japaridze, Edge dynamics in a quantum spin
  hall state: Effects from rashba spin-orbit interaction, Phys. Rev. Lett. 104
  (2010) 256804.

\bibitem{Budich12}
J.~C. Budich, F.~Dolcini, P.~Recher, B.~Trauzettel, Phonon-induced
  backscattering in helical edge states, Phys. Rev. Lett. 108 (2012) 086602.

\bibitem{Schmidt12}
T.~L. Schmidt, S.~Rachel, F.~von Oppen, L.~I. Glazman, Inelastic electron
  backscattering in a generic helical edge channel, Phys. Rev. Lett. 108 (2012)
  156402.

\bibitem{Oreg12}
N.~Lezmy, Y.~Oreg, M.~Berkooz, Single and multiparticle scattering in helical
  liquid with an impurity, Phys. Rev. B 85 (2012) 235304.

\bibitem{Crepin12}
F.~Cr\'epin, J.~C. Budich, F.~Dolcini, P.~Recher, B.~Trauzettel,
  Renormalization group approach for the scattering off a single rashba
  impurity in a helical liquid, Phys. Rev. B 86 (2012) 121106.

\bibitem{Konig12}
M.~K\"onig, M.~Baenninger, A.~G.~F. Garcia, N.~Harjee, B.~L. Pruitt, C.~Ames,
  P.~Leubner, C.~Br\"une, H.~Buhmann, L.~W. Molenkamp, D.~Goldhaber-Gordon,
  Spatially resolved study of backscattering in the quantum spin hall state,
  Phys. Rev. X 3 (2013) 021003.

\bibitem{Glazman13}
J.~I. V\"ayrynen, M.~Goldstein, L.~I. Glazman, Helical edge resistance
  introduced by charge puddles, Phys. Rev. Lett. 110 (2013) 216402.

\bibitem{Glazman14}
J.~I. V\"ayrynen, M.~Goldstein, Y.~Gefen, L.~I. Glazman, Resistance of helical
  edges formed in a semiconductor heterostructure, Phys. Rev. B 90 (2014)
  115309.

\bibitem{Pikulin14}
D.~I. Pikulin, T.~Hyart, Interplay of exciton condensation and the quantum spin
  hall effect in $\mathrm{InAs}/\mathrm{GaSb}$ bilayers, Phys. Rev. Lett. 112
  (2014) 176403.

\bibitem{Geissler14}
F.~Geissler, F.~Cr\'epin, B.~Trauzettel, Random rashba spin-orbit coupling at
  the quantum spin hall edge, Phys. Rev. B 89 (2014) 235136.

\bibitem{Kainaris14}
N.~Kainaris, I.~V. Gornyi, S.~T. Carr, A.~D. Mirlin, Conductivity of a generic
  helical liquid, Phys. Rev. B 90 (2014) 075118.

\bibitem{Fu09b}
L.~Fu, C.~L. Kane, Josephson current and noise at a
  superconductor/quantum-spin-hall-insulator/superconductor junction, Phys.
  Rev. B 79 (2009) 161408.

\bibitem{Beenakker12}
C.~W.~J. Beenakker, D.~I. Pikulin, T.~Hyart, H.~Schomerus, J.~P. Dahlhaus,
  Fermion-parity anomaly of the critical supercurrent in the quantum spin-hall
  effect, Phys. Rev. Lett. 110 (2013) 017003.

\bibitem{Crepin14a}
F.~Cr\'epin, B.~Trauzettel, Parity measurement in topological josephson
  junctions, Phys. Rev. Lett. 112 (2014) 077002.

\bibitem{Sticlet13}
D.~Sticlet, B.~D\'ora, J.~Cayssol, Persistent currents in dirac fermion rings,
  Phys. Rev. B 88 (2013) 205401.

\bibitem{Michetti11}
P.~Michetti, P.~Recher, Bound states and persistent currents in topological
  insulator rings, Phys. Rev. B 83 (2011) 125420.

\bibitem{Kane14b}
F.~Zhang, L.~Kane, C.\, Time-reversal-invariant ${Z}_{4}$ fractional josephson
  effect, Phys. Rev. Lett. 113 (2014) 036401.

\bibitem{Vondelft98}
J.~von Delft, H.~Schoeller, Annalen der Physik 7~(4) (1998) 225--305.

\bibitem{Giamarchi-book}
T.~Giamarchi, Quantum Physics in One Dimension, Oxford University Press, 2004.

\bibitem{GogolinBook}
A.~Gogolin, A.~Nersesyan, A.~Tsvelik, Bosonization and strongly correlated
  systems, Cambridge University Press, 2004.

\bibitem{Buttiker83}
M.~Büttiker, Y.~Imry, R.~Landauer, Josephson behavior in small normal
  one-dimensional rings, Physics Letters A 96~(7) (1983) 365 -- 367.

\bibitem{Buttiker85}
M.~B\"uttiker, Small normal-metal loop coupled to an electron reservoir, Phys.
  Rev. B 32 (1985) 1846--1849.

\bibitem{Buttiker85b}
R.~Landauer, M.~B\"uttiker, Resistance of small metallic loops, Phys. Rev.
  Lett. 54 (1985) 2049--2052.

\bibitem{Knez11}
I.~Knez, R.-R. Du, G.~Sullivan, Evidence for helical edge modes in inverted
  $\mathrm{InAs}/\mathrm{GaSb}$ quantum wells, Phys. Rev. Lett. 107 (2011)
  136603.

\bibitem{Knez14}
I.~Knez, C.~T. Rettner, S.-H. Yang, P.~Parkin, Stuart~S.\, L.~Du, R.-R. Du,
  G.~Sullivan, Observation of edge transport in the disordered regime of
  topologically insulating inasgasb quantum wells, Phys. Rev. Lett. 112 (2014)
  026602.

\bibitem{Maslov96}
D.~L. Maslov, M.~Stone, P.~M. Goldbart, D.~Loss, Josephson current and
  proximity effect in luttinger liquids, Phys. Rev. B 53 (1996) 1548--1557.

\bibitem{Fu06}
L.~Fu, C.~L. Kane, Time reversal polarization and a ${Z}_{2}$ adiabatic spin
  pump, Phys. Rev. B 74 (2006) 195312.

\end{thebibliography}

\end{document}